\documentclass[fleqn,usenatbib]{mnras}

\usepackage{newtxtext,newtxmath}

\usepackage[T1]{fontenc}
\usepackage{ae,aecompl}

\usepackage{graphicx}	
\usepackage{amsmath}	
\usepackage{amssymb}	
\usepackage[usenames,dvipsnames,svgnames,table]{xcolor}
\newcommand{\xmm}{XMM-\textit{Newton}}
\newcommand{\nustar}{NuSTAR}
\newcommand{\casa}{CASA \,}
\newcommand{\cloudy}{CLOUDY}
\newcommand{\cgs}{${\rm erg~cm}^{-2}~{\rm s}^{-1}$} 
\newcommand{\lum}{\rm erg~s$^{-1}$}

\title[Seyfert galaxy NGC~34]{CO excitation in the Seyfert galaxy NGC~34: stars, shock or AGN driven?}
\author[M. Mingozzi et al.]{
M. Mingozzi,$^{1,2}$\thanks{E-mail: matilde.mingozzi2@unibo.it}
L. Vallini,$^{3}$
F. Pozzi,$^{1}$
C. Vignali,$^{1,4}$
A. Mignano,$^{5}$
C. Gruppioni,$^{4}$
\newauthor{ M. Talia,$^{1}$
A. Cimatti,$^{1}$
G. Cresci,$^{2}$
M. Massardi$^{5}$}
\\
$^{1}$Dipartimento di Fisica e Astronomia, Università degli Studi di Bologna, Via Piero Gobetti 93/2, I-40129 - Bologna, Italy\\
$^{2}$INAF - Osservatorio Astrofisico di Arcetri, Largo E. Fermi 5, I-50157 - Firenze, Italy\\
$^{3}$NORDITA, KTH Royal Institute of Technology and Stockholm University, Roslagstullsbacken 23, SE-106 91 Stockholm, Sweden\\
$^{4}$INAF - Osservatorio Astronomico di Bologna, Via Piero Gobetti 93/2, I-40129 - Bologna, Italy\\
$^{5}$INAF - Osservatorio di Radioastronomia, Via Piero Gobetti 101, I-40129 Bologna, Italy
}
\date{Accepted 2017 November 18. Received 2017 November 18; in original form 2017 April 21}
\pubyear{2017}

\begin{document}
\label{firstpage}
\pagerange{\pageref{firstpage}--\pageref{lastpage}}
\maketitle

\begin{abstract}
We present a detailed analysis of the X-ray and molecular gas emission in the nearby galaxy NGC~34, to constrain the properties of molecular gas, and assess whether, and to what extent, the radiation produced by the accretion onto the central black hole affects the CO line emission. We analyse the CO Spectral Line Energy Distribution (SLED) as resulting mainly from Herschel and ALMA data, along with X-ray data from \nustar\ and \xmm. The X-ray data analysis suggests the presence of a heavily obscured AGN with an intrinsic luminosity of L$_{\rm{1-100\,keV}} \simeq 4.0\times10^{42}$~\lum. ALMA high resolution data ($\theta \simeq 0.2''$) allows us to scan the nuclear region down to a spatial scale of $\approx 100$ pc for the CO(6--5) transition. We model the observed SLED using Photo-Dissociation Region (PDR), X-ray-Dominated Region (XDR), and shock models, finding that a combination of a PDR and an XDR provides the best fit to the observations. The PDR component, characterized by gas density ${\rm log}(n/{\rm cm^{-3}})=2.5$ and temperature $T=30$ K, reproduces the low-J CO line luminosities. The XDR is instead characterised by a denser and warmer gas (${\rm log}(n/{\rm cm^{-3}})=4.5$, $T =65$ K), and is necessary to fit the high-J transitions. The addition of a third component to account for the presence of shocks has been also tested but does not improve the fit of the CO SLED. We conclude that the AGN contribution is significant in heating the molecular gas in NGC~34. 

\end{abstract}

\begin{keywords}
photodissociation region (PDR), galaxies: active, galaxies: ISM
\end{keywords}



\section{Introduction}\label{sez:intro}
Molecular gas is a key component of the interstellar medium (ISM), as it is the fuel of star formation and possibly of Active Galactic Nucleus (AGN) accretion. This gas phase is structured in gravitationally bound clouds (molecular clouds, MCs), characterised by a hierarchical structure that from the largest scale\footnote{In the solar neighborhood, the typical scale of MCs is $\approx 45\, \rm  pc$ \citep{GMCsMW}.} ($r\sim 10-30 \, $pc, $M \sim10^6 \, $M$_\odot$, $n$(H$_2)\sim10^2 $ cm$^{-3}$) extends down to small overdense regions ($r\approx 10^{-2} \,$pc, $M \sim 1 \, $M$_{\odot}$, $n$(H$_2)\sim10^6 \, $cm$^{-3}$), referred to as \textit{clumps} and \textit{cores} (e.g. \citealt{mckee_ostriker2007}). 

The molecular gas mass in galaxies is dominated by molecular hydrogen, H$_2$. Given the lack of a permanent dipole moment, the lowest rotovibrational transitions of molecular hydrogen are forbidden and have high-excitation requirements ($T_{\rm{ex}}\approx500$ K above the ground, significantly higher than kinetic temperatures in MCs, $T_{\rm{kin}}\sim15-100 \,$K). This is the reason why the molecular phase is generally traced through carbon monoxide (CO) rotational transitions (e.g., \citealt{mckee_ostriker2007, carilli}) being the CO the second most abundant molecule after H$_2$ in the Universe \citep[${\rm CO/H_2} \approx 1.5 \times 10^{-4}$ at solar metallicity][]{lee96}.
 
More importantly, the study of the so called CO Spectral Line Energy Distribution (CO SLED) -- i.e. the CO line luminosity as a function
of the CO upper rotational level -- allows us to constrain the physical properties of the molecular gas, both at high-redshift ($z > 1$)(e.g.,
\citealt{weiss07, gallerani14, daddi15}), thanks to sub-millimeter facilities such as CARMA and IRAM-PdBI interferometers, and in local sources 
(e.g., \citealt{rosenberg, mashian15, lu17, pereirangc34, pozzi17}), with Herschel. In the last few years, the ALMA advent has represented a 
breakthrough making it possible to reach a spatial resolution of $50-100$~pc when targeting CO lines in the nearby Universe (e.g., 
\citealt{xu14,xu2015,zhang2016}.) 

The critical densities of the CO rotational transitions rise from $n_{\mathrm{crit}} \simeq 10^3 \, \rm cm^{-3}$ for the CO(1--0) to $n_{\mathrm{crit}} \simeq 10^6 \, \rm cm^{-3}$ for $J_{\mathrm{up}}=13$ \citep{carilli}. 
This makes the CO(1--0) the most sensitive transition to the total gas reservoir, being excited in the diffuse cold ($n_{\mathrm{crit}} \simeq 2.1 \times 10^3 \rm \, cm^{-3}$, $T_{\rm{ex}} \simeq 5.5\, \rm K$) molecular component. 
The higher-J transitions ($J_{\mathrm{up}}>1$) are, instead, increasingly luminous in the warm dense ($n_{\mathrm{crit}} \simeq 0.01 - 2 \times 10^6\rm \,cm^{-3}$, $T_{\rm{ex}} \simeq 16.6 - 7000 \rm \,K$) star-forming 
regions within MCs \citep{carilli}. 
The CO SLED shape is determined by (i) the gas density  and (ii) the heating mechanism acting on the molecular gas. The heating agents can be either far-ultraviolet (FUV, $6-13.6$~eV) photons, X-ray photons, 
or shocks \citep[e.g.][]{COexcitation}.
Regions whose molecular physics and chemistry are dominated by FUV radiation, e.g. from OB stars, are referred to as \textit{Photo-Dissociation Regions} (PDRs, \citealt{holl99}), while those influenced by X-ray photons 
($1-100$~keV), possibly produced by an AGN, are referred to as \textit{X-ray-Dominated Regions} (XDRs, \citealt{mal96}). Interstellar shock waves, instead, originate from the supersonic injection of mass into the ISM by 
stellar winds, supernovae, and/or young stellar objects, that compress and heat the gas, producing primarily H$_2$, CO, and H$_2$O line emission, and relatively weak dust continuum \citep{holl_shocks}.
In PDRs, the CO emission saturates at a typical column densities of ${\rm log}(N_{\rm{H}}/{\rm cm^{-2}})\approx 22$ \citep{holl99} and the resulting CO SLED considerably drops and flattens at high-J transitions 
\citep{vandishoeck}.
On the contrary, the smaller cross sections of X-ray photons allow them to penetrate deeper than FUV photons\footnote{A $1\rm  \,keV$ photon penetrates up to a typical column of $N_{\rm{H}}\simeq 2 \times 
10^{22}$~cm$^{-2}$, whereas a 10 keV photon penetrates $N_{\rm{H}}\simeq 4 \times 10^{25}$~cm$^{-2}$, \citep{schleicher10}}. Because of that, the XDRs are characterised by column densities of warm molecular gas, where 
high-J CO can be efficiently excited. The resulting CO SLEDs are thus peaked at higher-J rotational transitions compared to those resulting from PDRs.
The same holds true for shock-heated molecular gas.
Shocks can heat the gas above $T\approx 100\rm \, K$, and at such high temperatures the high-J CO rotational energy levels become more populated and the resulting CO SLED peaked at high-J.

The goal of this paper is to shed light on the ``ambiguous" nature of the local galaxy NGC~34, exploiting the peculiarities of its CO SLED. NGC~34 has been either classified as a starburst galaxy, because of the lack of high-ionisation emission lines, such as the [Ne V] ($12.3\rm \, \mu m$ and $24.3 \rm \, \mu m$) and the [O IV] ($25.89 \rm  \,\mu m$), in the infrared (IR) spectrum, and as a Seyfert 2 galaxy, owing to the optical spectrum properties and X-ray observations.

We aim at assessing whether and to what extent the CO emission is influenced by the radiation produced by the accretion onto the black hole. To do that we analyse the X-ray and CO emission, using mainly archival data from \xmm, \nustar, ALMA, and Herschel.
On the one hand, the X-ray data allow us to properly include the effect of AGN radiation in the modelling of the CO SLED, on the other hand, the CO emission, traced by ALMA in the central region of NGC~34 at high spatial resolution, is crucial to spatially constrain the region where the contribution of the AGN actually dominates.

The paper is structured as follows. In Sec.~\ref{sez:2}, the observations, data reduction and line luminosities are discussed, while in Sec.~\ref{sez:3} we report the model assumptions, the results and the discussion. Finally, in Sec.~\ref{sez:4} we conclude with a summary of the results. We took into account a $\Lambda CDM$ cosmology with $H_0=69.9$ km s$^{-1}$ Mpc$^{-1}$, $\Omega_M=0.29$ and $\Omega_{\Lambda}=0.71$, that yields a luminosity distance D$_L$=85.7 Mpc at $z \simeq0.0196$, with 1'' corresponding to 0.4 kpc.

\section{Data}\label{sez:2}

\subsection{NGC~34: a nearby ambiguous object}\label{sez:ngc34}
NGC~34 is a local object ($z\simeq0.0196$, $2.2'\times0.8'$), classified as a Luminous Infrared Galaxy (LIRG; ${\rm log} (L_{{\rm 8-1000\,\mu m}})\simeq11.44 \rm \, L_\odot$, \citealt{sanders2003}).

\citet{riffel06} studied NGC~34 near-infrared (NIR; $0.8-2.4 \rm \,\mu m$) spectrum, that appears to be dominated more by stellar absorption features rather than emission lines (e.g., [S III], typical of Seyfert 2 spectra), suggesting that this source is not a ``genuine'' AGN or that it has a buried nuclear activity at a level that is not observed at NIR wavelengths. Additional support to this conclusion comes from the lack of high-ionisation lines in the Spitzer-IRS spectrum, such as the [Ne V] ($12.3 \rm \, \mu m$ and $24.3\rm \, \mu m$) and the [O IV] ($25.89 \, \rm \mu m$), that are exclusively excited by AGN, and can be considered AGN spectral signatures \citep{tommasin10}. Consequently, \citet{riffel06} infer that NGC~34 is a starburst galaxy.

\citet{mulchaey1996} analyse the optical emission lines of NGC~34, inferring that it is a rather weak emitter 
of [OIII]$\lambda$5007, when compared to other Seyfert galaxies in their sample, while it shows a strong H$
\alpha$ emission. Therefore, they conclude that the ionisation of the gas is not related to any Seyfert activity. Nevertheless, a later work by \citet{brightman11_2} reports new optical emission line ratios, confirming the presence 
of a Seyfert 2 nucleus through optical diagnostic diagrams (BPT, \citealt{baldwin81}). \citet{vardoulaki15} confirmed the presence of an AGN, analysing the radio spectral index maps.  Furthermore, NGC~34 appears to have an 
intrinsic X-ray luminosity of ${\rm log}(L_\mathrm{2-10\,keV})\simeq 42 \rm \, erg\, s^{-1}$ on the basis of \xmm \,data \citep{brightman11}. Assuming the Ranalli relation \citep{ranalli}, this large X-ray luminosity can be 
explained only by a $SFR>100 \rm \, M_{\odot}\,yr^{-1}$, while NGC~34 SFR is $\simeq24\,$M$_{\odot}\,$yr$^{-1}$ \citep{gruppioni}. 
Another indication of the presence of an AGN has been reported by \citet{gruppioni}, who carried out a SED decomposition analysis, suggesting the presence of circumnuclear dust (i.e., dusty torus component) heated by a 
central AGN. Moreover, the mid-infrared (MIR; $5-20 \rm \,\mu m$) intrinsic luminosity corresponding to the torus appears to correlate with the X-ray luminosity in the 2--10 keV band, reinforcing the hypothesis of an AGN 
\citep{gandhi09}. 

\subsection{X-ray data}\label{sez:xray}
NGC~34 was observed by \xmm\ on Dec. 22$^{nd}$, 2002. After cleaning the data to remove periods of high background, the cleaned exposure time was 9.87~ks in the pn camera and 17.12~ks and 16.75~ks in the MOS1 and MOS2 cameras, respectively. The source spectra were extracted from circular regions of radius 30\arcsec\ for the pn and 18\arcsec\ for the MOS cameras, centered on the optical position of the source; background spectra were extracted from larger regions, free of contaminating sources, in the same CCD as the source spectra. The number of net (i.e., background-subtracted) counts is 670 (pn), 280 (MOS1) and 240 (MOS2). Then the spectra from the two MOS cameras were summed using {\sc mathpha} to increase the statistics in each spectral bin, along with the corresponding background and response matrices. pn (MOS1$+$2) spectra were binned to 20 (15) counts/bin to apply $\chi^2$ statistics; in the following X-ray spectral analysis (carried out using {\sc xspec}; \citealt{arnaud}), errors and upper limits are reported at the 90\% confidence level for one parameter of interest \citep{avni}. 

\nustar\ observed NGC~34 on July 31$^{st}$, 2015, for an exposure time of 21.2~ks. Data were reprocessed and screened using standard settings and the {\sc nupipeline} task, and source and background spectra (plus the corresponding response matrices) were extracted using {\sc nuproducts}. A  circular extraction region of 45\arcsec\ radius was selected, while background spectra were extracted from a nearby circular region of the same size. The source net counts, $\approx$130 and $\approx$120 in the FPMA and FPMB cameras, respectively, were rebinned to 20 counts per bin. The source signal is detected up to $\approx$25~keV. 

\begin{figure}
 \centering
 \includegraphics[width=1.1\columnwidth]{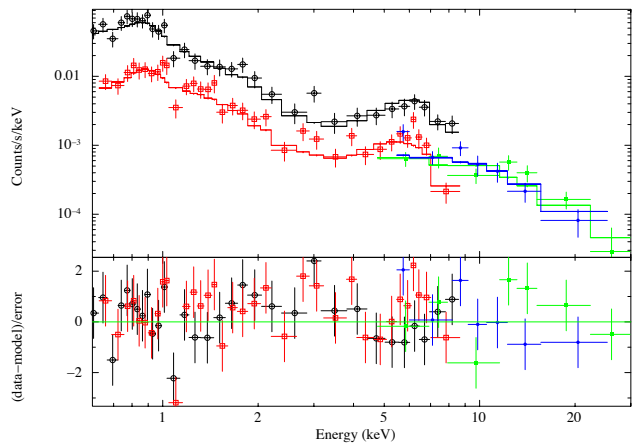}
 \caption{Top panel: \xmm\ pn and MOS (black and red points, respectively) and \nustar\ FPMA and FPMB 
(green and blue points, respectively) spectral data and BF modelling (continuous lines); see text for details. Bottom panel: spectral residuals in units of $\sigma$.}
 \label{fig:x_spec}
\end{figure}

At first we verified that no large flux variability occurs between \xmm\ and \nustar\ data, being well within a factor of two. Then the four datasets were fitted simultaneously with the same modelling. The advantage of using all datasets is to achieve a proper source modelling over a broad energy range. To this goal, we adopted a model consisting of an absorbed power-law, accounting for the continuum emission above a few keV, and a thermal plus scattering component below 2 keV. Due to the limited photon statistics, we fixed the power-law photon indices (of the nuclear and scattered components) to $\Gamma=1.9$, as typically found in AGN and quasars (e.g., \citealt{piconcelli}). For the thermal emission, likely related to the host galaxy, we used the {\sc mekal} model, obtaining kT=0.7$\pm{0.1}$~keV. The obscuration towards the source is $N_{\rm{H}}=5.2^{+1.3}_{-1.1}\times 10^{23}\,$cm$^{-2}$, consistent with previous analyses of the \xmm\ datasets (\citealt{guainazzi05}; \citealt{brightman11}) and the recent one conducted on \xmm\ and \nustar\ data by \citet{ricci17}. About 4\% of the nuclear emission is scattered at soft X-ray energies. An upper limit to the iron K$\alpha$ emission line of 110~eV is derived. The best-fitting model and data$-$model residuals are reported in Fig.~\ref{fig:x_spec}. The observed 2--10~keV flux is 3.2 $ \times 10^{-13}$~\cgs, while the intrinsic (i.e., corrected for the obscuration) rest-frame 2--10~keV and 1--100~keV luminosities are 1.3 $ \times 10^{42}$~\lum\ and 4.0 $ \times10^{42}$~\lum, respectively. Given that, the AGN appears to contribute to a 10\% to the bolometric luminosity of the galaxy.

\subsection{ALMA data}\label{sez:alma}
NGC~34 nuclear region was observed with ALMA in Band 9 during the Early Science Cycle 0 in May and August 2012 (project 2011.0.00182.S, PI: Xu), targeting the CO(6--5) transition ($\nu_{\mathrm{rest-frame}}=$~691.473 GHz) and the 435~$\mu$m dust continuum. 
The observing time is split in five Execution Blocks (EBs) with 16 antennas, set up in an extended configuration, and in one EB with 27 antennas set up in both a compact and an extended configuration (the minimum baseline is 21~m, the maximum baseline is 402~m). Hence, at the observing band, the largest angular scale (LAS) recovered by ALMA is $\approx 4$\textacutedbl, corresponding to $\sim2$~kpc at NGC~34 distance, while the spatial resolution of the observation is $\theta \simeq 0.2$\textacutedbl, corresponding to $\sim100$~pc. The field of view for the used 12 m antennas is FOV~$\approx 9$\textacutedbl. The medium precipitable water vapour of all the observing blocks is PWV $\simeq0.38$ mm. 
The observation was performed in time division mode with a shallow velocity resolution of 6.8 km s$^{-1}$, with four spectral windows (SPWs) of $\approx2 \,$~GHz, centered at the sky frequencies of 679.8, 678.0, 676.3 and 674.3 GHz, respectively.  The on-source integration time is about 2.25 hr. The phase, bandpass and flux calibrators observed are 2348-165, 3C 454.3 and the asteroid Pallas, respectively, and are the same for all the EBs. The analysis of the data obtained with a standard pipeline was reported by \citet{xu14}.

Cycle 0 data re-processing is strongly recommended, since, thanks to the new flux model libraries available (\textit{Butler-JPL-Horizons 2012}), a more reliable flux calibration can be obtained \citep{almamemo594}\footnote{\citet{almamemo594} contend that many of the models used to calculate the flux density of solar system bodies in the 2010 version were inaccurate with respect to the new ones of 2012, that use new brightness temperature models and a new flux calculation code that replaces the ``Butler-JPL-Horizons 2010'' standard used before.}. Therefore, we took the raw data from the archive, we generated new reduction scripts and we calibrated the data using \casa4.5.2, correcting for antenna positions, atmosphere effects and electronics.
Since these data are Band 9 observations, the calibrators are faint and the signal-to-noise ratio of the solutions is low, so that a phase solution over 60 s and a bandpass solution over 30 channels can be obtained.
In addition, we put the signal-to-noise threshold equal to 1.5 (3 would be the recommended value) to find as many solutions as possible. Then, we applied the more reliable flux calibration, obtaining, for the calibrators, a $\approx 20 \%$ higher flux density than the one reported in the archive.

We applied the self calibration, making the image of the emission by selecting only the channels where the emission line was detected, we built a model and we found phase solutions averaged on 600 s (i.e., approximately two scans). Then, we iterated again and we found phase and amplitude solutions, averaging 1200 s and combining scans. We produced the channel map of the CO(6--5) transition with a rms of $\approx6.5 \,$mJy/beam, subtracting the continuum, binning into channels with a width of $\approx 34$ kms$^{-1}$, in order to increase the signal-to-noise ratio, and iteratively cleaning the dirty image, selecting a natural weighting scheme of the visibilities.  

\begin{figure}
\centering
\includegraphics[width=1.\columnwidth]{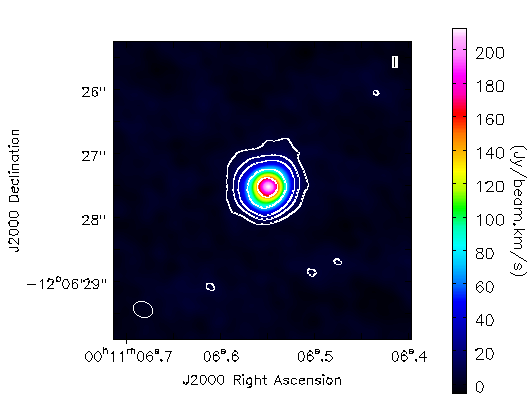}
\caption{Integrated emission of the CO(6--5) line. The wedge on the right shows the color-scale of the map in Jy beam$^{-1}$ km s$^{-1}$. The integrated flux density is $(707\pm106)$ Jy km s$^{-1}$. The contours at [3,10,20,50,100] are set in units of the rms noise level of the image, rms $\approx1.5$ Jy beam$^{-1}$ km s$^{-1}$.}
\label{fig:mom0}
\end{figure}

Fig. \ref{fig:mom0} shows the integrated emission of the CO(6--5) transition. Considering the emission within 3$\sigma$, contained in a diameter of $\theta \approx$ 1.2\textacutedbl ($\approx 500$ pc), we obtain an integrated flux of $(707 \pm 106) \, \mathrm{Jy \, km \, s^{-1}}$ (i.e., $(1.63\pm0.24)\times10^{-14}\,$erg$ \, $s$^{-1} \, $cm$^{-2}$) with a peak of $(213 \pm 32) \, \mathrm{Jy \, km \, s^{-1}}$. The rms of the image is $\approx1.5$ Jy/beam km s$^{-1}$. The errors are totally dominated by the calibration error, that is the 15\% of the total flux in Band 9 for Cycle 0 Data. Our integrated flux is $\approx30\%$ lower than the one reported by \citet{xu14}. The cause of this discrepancy is likely due to the different flux calibration and the use of the new version of the \casa software in our analysis.

\subsection{CO data}\label{sez:co}
\begin{figure}
 \centering
 \includegraphics[width=1.\columnwidth]{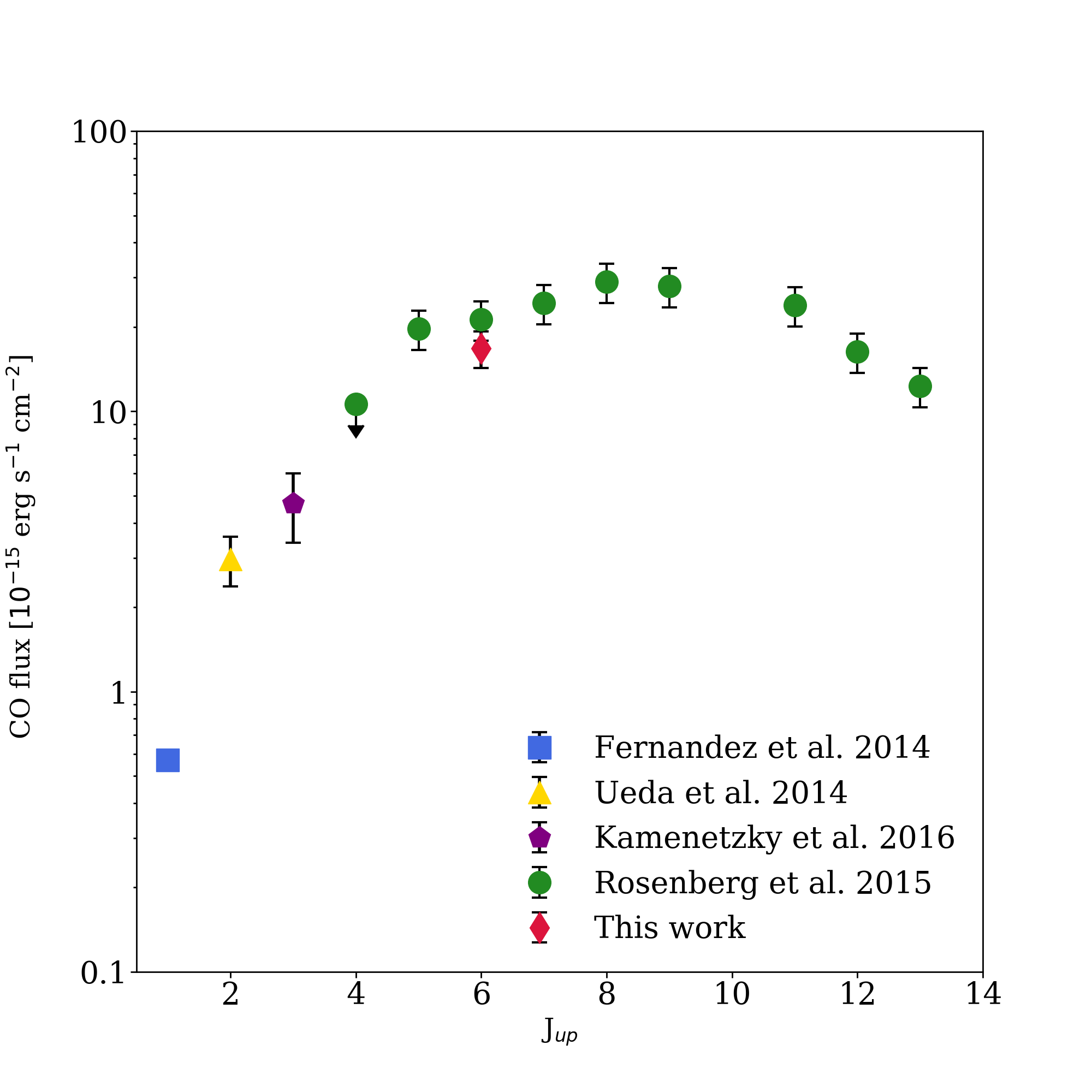}
 \caption{Observed CO flux of NGC~34 as a function of the upper J rotational level. The blue square, yellow triangle and purple pentagon represent the ground-based observations (\citealt{fernandez,ueda14,kamenetzky16}), the red diamond the CO(6--5) transition observed with ALMA and the green circles the Herschel/SPIRE FTS data from $J_{\mathrm{up}}=4$ to $J_{\mathrm{up}}=13$ \citep{rosenberg}.}
 \label{fig:cosled_obs}
\end{figure}

NGC~34 CO SLED is shown in Fig.~\ref{fig:cosled_obs}, where the CO rotational transitions up to $J_{\mathrm{up}}=13$ are reported. The low-J transitions are ground-based observations. The CO(1--0) transition (blue square) was observed by CARMA, that covered a FOV $\approx 1.7'$, detecting a regularly rotating disk of molecular gas with a diameter of $2.1 \,$kpc \citep{fernandez}. The CO(2--1) transition (yellow triangle) was observed by the Submillimeter Array (SMA) with a FOV $\approx 22$\textacutedbl \citep{ueda14}. By comparing the observed flux with the single-dish measurement made with the Swedish ESO Submillimeter Telescope (SEST), that covered a FOV $\approx 1'$ \citep{albrecht}, we note that SMA recovered all the flux \citep{ueda14}. The observation of the CO(3--2) transition (purple pentagon) was conducted with the Submillimeter Telescope (SMT) with a FOV $\approx 22$\textacutedbl  and the measure was corrected for subsequent comparison to Herschel CO lines \citep{kamenetzky16}.
The CO(6--5) transition (red diamond) was observed by ALMA and re-calibrated through the procedure outlined in Sec.~\ref{sez:alma}, while the transitions from $J_{\mathrm{up}}=4$ to $J_{\mathrm{up}}=13$ (green circles) are Herschel/SPIRE observations, sparsely covering a FOV~$\approx 2$' \citep{rosenberg}. NGC~34 appears point-like in the Herschel/SPIRE photometric bands (from 250 to 500 $\mu$m), characterised by a beam size of 17\textacutedbl$-$ 42\textacutedbl ($\approx7-20\rm \, kpc$). Assuming that the dust (sampled by the SPIRE photometric observations) and the gas (sampled by SPIRE/FTS observations) are almost co-spatial, it can be argued that all the CO fluxes correspond to the integrated emission of the galaxy.  
By comparing the two different data sets for the CO(6--5) transition, we find a good agreement within 1$\sigma$. 
This allows us to be confident that the ALMA observation has recovered all the flux, which means that no flux related to the CO(6--5) emission comes from a region larger than the one observed with ALMA. Moreover, since ALMA LAS is comparable in size to the CO(1--0) disk (LAS $\approx 2$ kpc, see Sec.~\ref{sez:alma}), we can assume that no flux has been lost on the largest scales. Hence, all the data taken from the literature map the source entirely and are thus comparable in size.

\section{Interpreting the CO SLED of NGC~34}\label{sez:3}
To fit and interpret the CO SLED of NGC~34, we have modelled the effect of the FUV and X-ray radiation on the molecular gas with the version c13.03 of \cloudy~\citep{cloudy}, and the effect of the mechanical heating by adopting the shock models of \citet{flower2015}. In what follows we separately discuss their main features.

\subsection{PDR and XDR modelling}
We run a grid of \cloudy \, PDR and XDR models that span ranges in density ($n$), distance between the source and the illuminated slab of the cloud ($d$)\footnote{A larger distance from the source implies a lower incident radiation field, since the flux is proportional to d$^{-2}$. Hence, taking into account different values of distance means to vary the incident radiation field.}, and column density ($N_{\rm{H}}$). 
In these models, the external radiation field impinges the illuminated face of the cloud, that is assumed to be a 1-D gas slab with a constant density, at a fixed distance from the central source (i.e., the collective light of stars in PDRs and an X-ray point-like source in XDRs).

For the PDR models the SED of the stellar component is obtained using the stellar population synthesis code Starburst99, assuming a Salpeter Initial Mass Function in the range 1-100 $M_{\odot}$, Lejeune-Schmutz stellar 
atmospheres (\citealt{schmutz}, \citealt{lejeune}), solar metallicity and a continuous star formation mode, that is normalised to NGC~34 SFR ($\simeq24\,$M$_{\odot}\,$yr$^{-1}$, \citealt{gruppioni}). We assume a Milky Way gas-to-dust ratio of $\approx160$ (e.g., \citealt{dustgasratio}) for the gas slab.
We run 21 PDR models, varying $n$ in the range ${\rm log}$(n/cm$^{-3})=[2-4.5]$ and the distance of the gas slab from the radiation source in the range $d= [100-500]$ pc. These values of distances translate into a flux in the 
FUV band normalised to that observed in the solar neighbourhood ($1.6 \times 10^{-3}\, \rm{erg\, s^{-1}\, cm^{-2}}$, \citealt{habing68}), in the range $[10^4-10^3]$ G$_0$, respectively. The code computes the radiative 
transfer through the slab spanning a hydrogen column density range ${\rm log}$($N_{\rm{H}}$/cm$^{-2})=[19-23]$. The densities and column densities are chosen to cover the typical values in giant MCs \citep{mckee}. We 
constrain the range of distances to be larger than the ALMA minimum recoverable scale ($\approx100$ pc, see Sec.~\ref{sez:alma}). 

In the XDR models, the AGN radiation field is modelled following the default \texttt{table AGN} \cloudy command \citep{korista}, and its spectrum is normalised so that the 1-100 keV X-ray luminosity matches the observed one $L_{\mathrm{1-100\,keV}}\simeq 4 \times 10^{42}$ erg s$^{-1}$ (see Sec.~\ref{sez:xray}). 
We run 9 models, adopting a slightly higher range of densities ${\rm log}$(n/cm$^{-3})=[3.5-5.5]$ and the same values of column densities and distances, that translate into an X-ray flux at the cloud surface F$_X\simeq 2.2-0.2$ erg$\,$cm$^{-2}\,$s$^{-1}$.

\subsection{Shock modelling}
The \citet{flower2015} shock code simulates shocks in the ISM as a function of the physical conditions in the ambient gas, providing the CO line intensities (i.e. fluxes per unit of the emitting area where the shock acts, $A$). 
 Shock waves in the ISM can be distinguished in ``jump" (J) or ``continuous" (C) types, according to the strength of the magnetic field and the ionisation degree of the medium in which they propagate \citep[see e.g. Fig. 1 in][]{draine80}. In J-shocks the variation in the fluid properties (density, temperature, and velocity) occurs sharply and can be approximated as a discontinuity. In C-shock models, instead, the influence of the magnetic field on the ionised gas causes the ions to diffuse upstream of the shock front, thus eliminating the discontinuities in the fluid variables \citep[][]{draine93,pon2012}. These differences cause the J-shock temperature (T$\approx 10^4$~K) to be far higher than the one characterising C-shock (T$\approx 10^3$~K), inducing the dissociation of H$_2$ molecules \citep{flower2010}. This is the reason why C-type shocks are preferentially invoked to affect the CO emission in molecular clouds (e.g., \citealt{smith97}, \citealt{hailey2012}).

In the present work, we compare the observed CO SLED with two grids of C- and J-shock models, respectively. We let the shock velocity\footnote{Shock velocities greater than 50 km s$^{-1}$ are known to be fast enough
to destroy molecules \citep{holl_shocks}.} vary in the range $v_{sh}=[10-40] \rm \, \, km\, s^{-1}$, and the pre-shock density in ${\rm log}(n/{\rm cm^{-3}})=[3-6]$. The transverse magnetic field scales with the density through the relation $B \sim b {\rm n^{1/2}}$, where $b=1$ for C-shocks, and $b=0.1$ for J-shocks \citep{flower2015}.

\subsection{CO SLED fitting}\label{sez:fitting}
\begin{figure}
 \begin{minipage}[b]{.50\textwidth}
  \centering
  \includegraphics[width=1.\columnwidth]{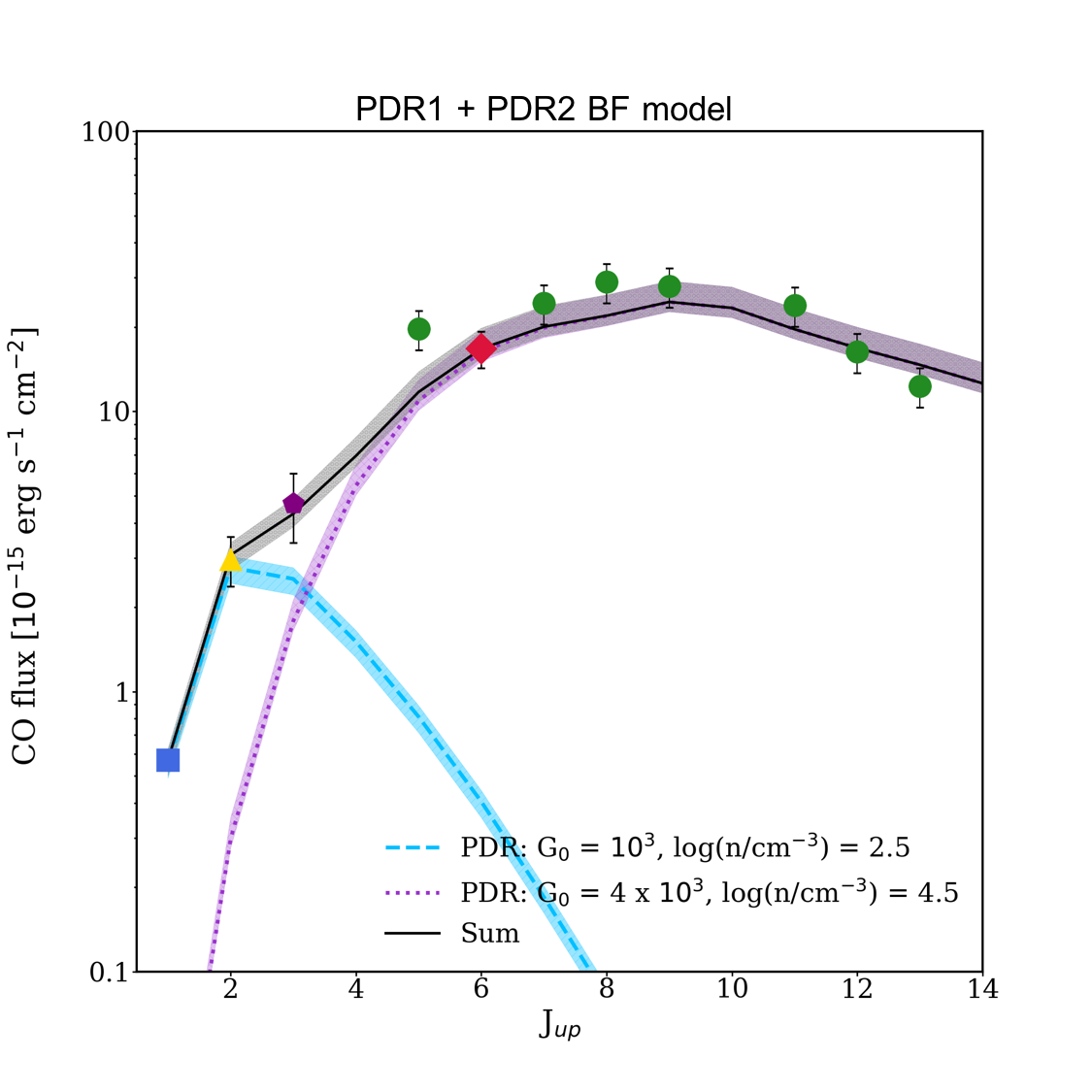}
 \end{minipage}
 \hspace{0.3cm} 
\begin{minipage}[b]{.50\textwidth}
  \centering
  \includegraphics[width=1.\columnwidth]{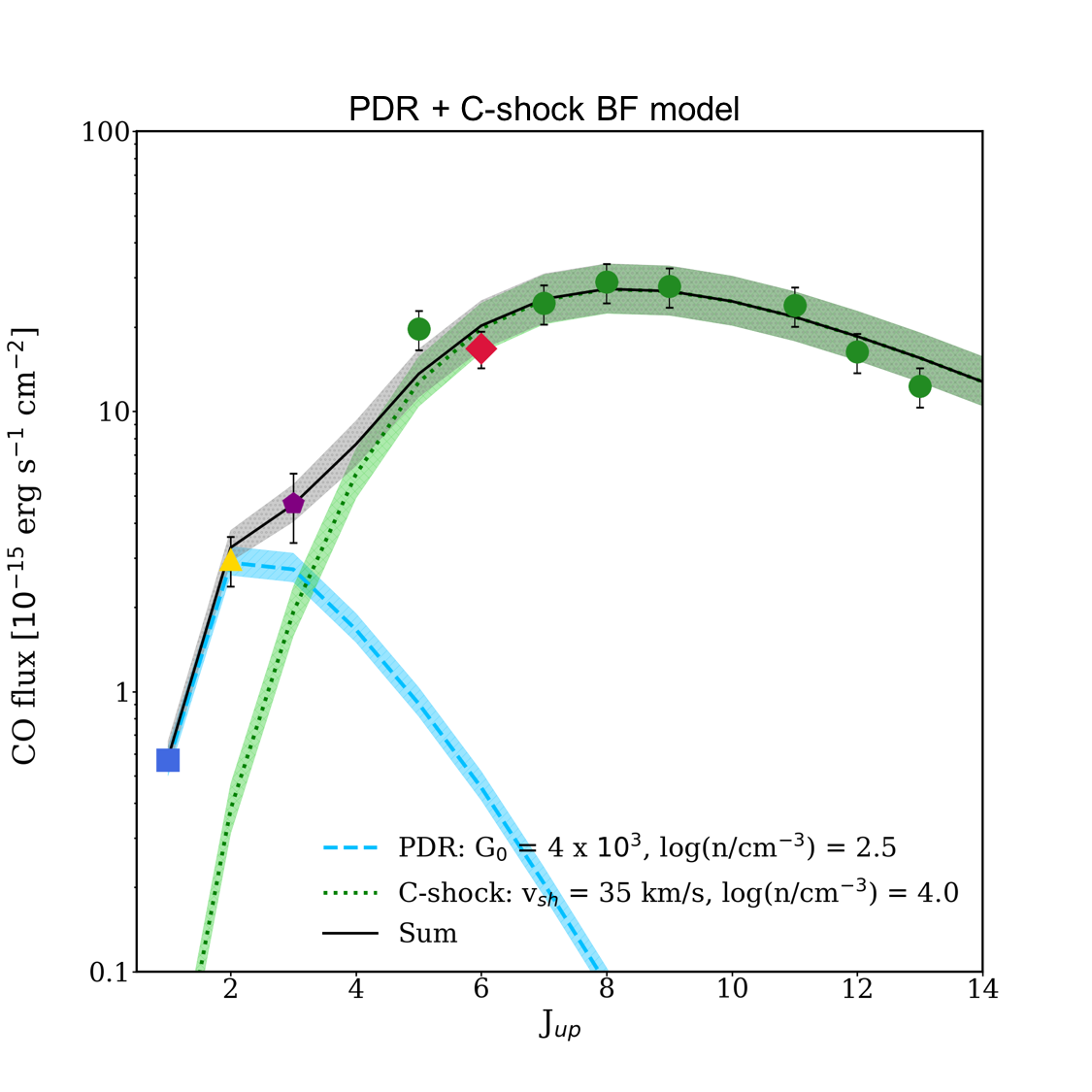}
 \end{minipage}
 \caption{Top panel: PDR1+PDR2 BF model, overplotted to the observed data. The light-blue dashed line and the purple dotted line represent the low-density and high-density PDR, respectively. Low panel: PDR+C-shock BF model, overplotted to the observed data. The light-blue dashed line and the green dotted line indicate the low-density PDR and the C-shock component, respectively. In these figures, the black solid line indicates the sum of the two components and the shaded areas indicate the $\pm 1 \sigma$ uncertainty range on the normalisation of each component. }
 \label{fig:pdrshockmodels}
\end{figure}

\begin{figure*}
 \centering
 \includegraphics[width=1.2\columnwidth]{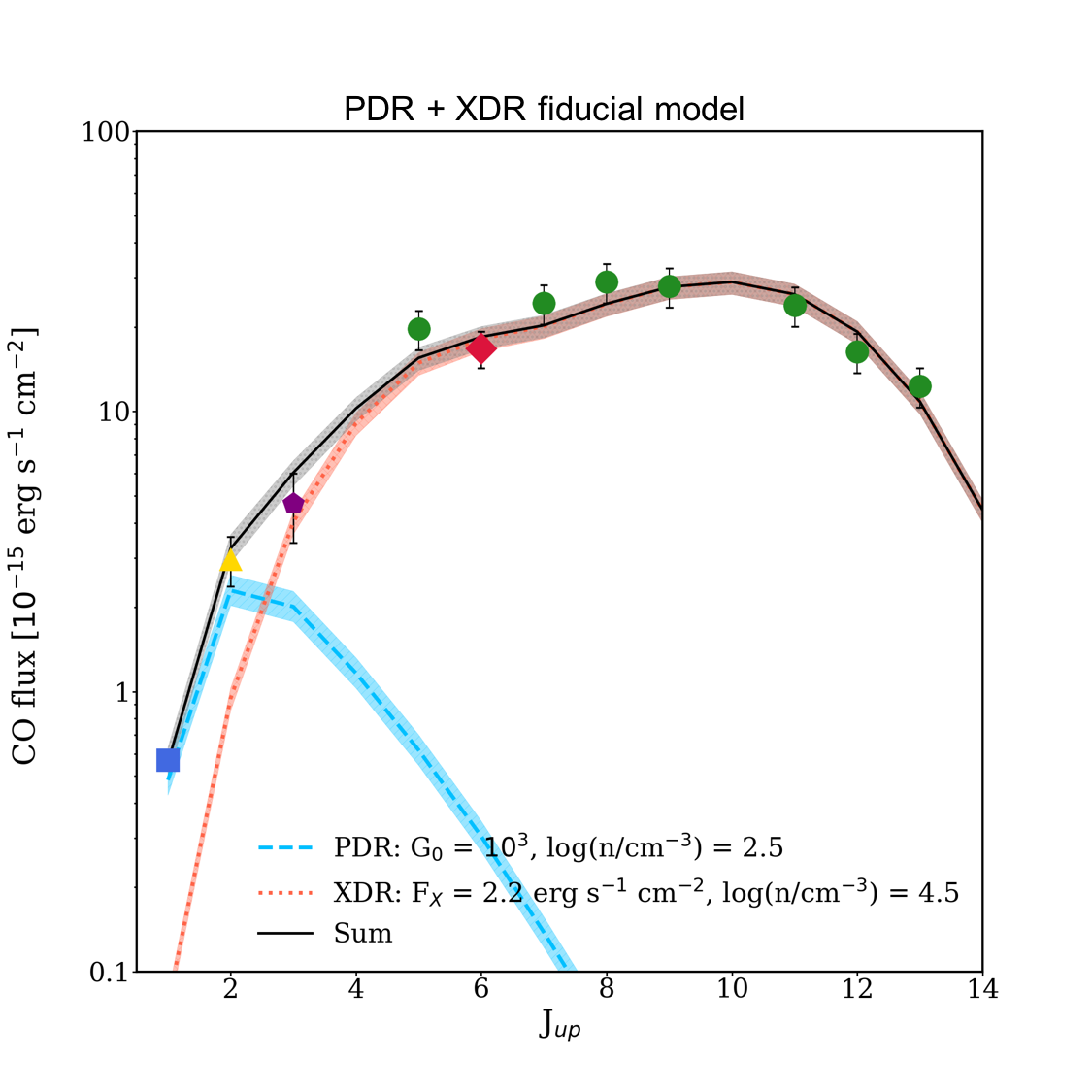}
 \caption{Fiducial model (PDR+XDR), overplotted to the observed data. The light-blue dashed line and the red dotted line represent the low-density PDR and the high-density XDR, respectively. The black solid line indicates the sum of the two components and the shaded areas indicate the $\pm 1 \sigma$ uncertainty range on the normalisation of each component.}
 \label{fig:best-fit3}
\end{figure*}

Despite the good spatial resolution of ALMA (see Sec. \ref{sez:alma}), individual clouds in extragalactic sources cannot be resolved. Each resolution element in ALMA observations measures the combined emission from a large ensemble of molecular clouds.
Moreover, as pointed out in Sec. \ref{sez:intro}, low-J CO transitions are generally excited in the cold diffuse component of molecular gas while mid- and high-J transitions are associated to a warmer and denser component. 
The observed CO SLED of NGC 34 can then be reproduced by summing a low-density PDR component (e.g., ${\rm log}(n$/cm$^{-3})\simeq 2-2.5$) that accounts for the low-J transitions, to a second component for the high-J transitions. 
Given the spatial constraints on the CO(1--0) emission (see Sec.~\ref{sez:co}), the low-density PDR must account for the emission of MCs located within $d< 1$ kpc from the centre of the galaxy. 
In addition, we fixed the gas column density of the low-density PDR to ${\rm log}$($N_{\rm{H}}$/cm$^{-2})\simeq 21.8$, that is slightly above that (${\rm log}$($N_{\rm{H}}$/cm$^{-2})\simeq 21.6$) required to have CO molecules \citep{vandishoeck}. The CO(6--5) emission line allows to constrain the high-density component in a region of $d<250$ pc from the centre.
We have adopted three different approaches to model this second component, that can be either produced by:
\begin{itemize}
\item a high-density PDR (PDR1+PDR2 model);
\item the shock heated gas (PDR+shock model);
\item an XDR (PDR+XDR model). 
\end{itemize}
The PDR1+PDR2 and the PDR+shock models are shown in Fig.~\ref{fig:pdrshockmodels}, in the upper and lower panels, respectively. The black solid line represents the sum of the components, while the 1$\sigma$ confidence levels on the normalisations have been obtained by marginalising over the other parameters and considering a $\Delta \chi^2 = 2.3$ \citep{lampton76}. The PDR+XDR model is shown in Fig.~\ref{fig:best-fit3}. All the parameters taken into account and their BF values are summarised in Tab. \ref{tab:tabbest-fit}.

\begin{table}
	\centering
	\caption{PDR1+PDR2 and PDR+C-shock BF models, and PDR+XDR fiducial model parameters. For PDRs and XDRs, the distance from the source, the density and the column density values are reported. Density and column density values are given in logarithm. For C-shocks, the shock velocity, the pre-shock density and the shock-related emitting area are reported. The * means that the value was fixed in the chi-square analysis. Finally, the ${\chi}^2$ and degrees of freedom (dof) values are indicated.}
	\label{tab:tabbest-fit}
\begin{tabular}{lccccccccr}  		
		\hline
		  Best-fit  	 	  &   d             & n                          & N                         	\\
					  & $[$pc$]$   & $[$cm$^{-3}]$      & $[$cm$^{-2}]$  	 			\\		 
		 \hline
		PDR1          & 500   	     & 2.5    		   &   21.8*		 	 \\
		PDR2            & 250	     & 4.5   			   &   21.7  		   			 \\
		\hline
		$\chi^2$/dof &		   &13.2/4 		 	&\\
		\hline
		\hline
		  Best-fit  	 	  &   d             & n                          & N                         	\\
					  & $[$pc$]$   & $[$cm$^{-3}]$      & $[$cm$^{-2}]$  	  		\\		 
		 \hline
		PDR                   & 500   	     & 2.5    		   &   21.8*		 	                         \\
		\hline
		 	 	          &   v$_{sh}$ & n                          & A                         	                \\
					 & $[$km s$^{-1}]$   & $[$cm$^{-3}]$      & $[$pc$^{2}]$   	       		\\	
		\hline
		C-shock             & 35 	     & 4.0  			   &   $($785$)^2$              			 \\
		\hline
		$\chi^2$/dof	&&9.9/5&\\
		\hline
		\hline
		  Best-fit  	 	  &   d             & n                          & N                          		\\
					  & $[$pc$]$   & $[$cm$^{-3}]$      & $[$cm$^{-2}]$  	    		 		\\		 
		 \hline
		PDR                   & 500   	     & 2.5    		   &   21.8*		  \\
		XDR                   & 100 	     & 4.5   			   &   23.0*  	 			 \\
		\hline
		$\chi^2$/dof     &&7.8/5&\\
		\hline
	\end{tabular}
\end{table}

The PDR1+PDR2 best-fit (BF) model is characterised by a value of the reduced-${\chi}^2$ ($\tilde{\chi}^2$), defined as the ratio between the computed ${\chi}^2$ and the related degrees of freedom, of $\tilde{\chi}^2_{BF}= 3.3$. Such a high value made us infer that the high-density PDR cannot reproduce completely the high-J transitions.

The PDR+C-shock BF model has instead a slightly lower $\tilde{\chi}^2_{BF}$ ($= 2.0$). However, the best-fit for the C-shock-heated component, accounting for the high-J CO emission, returns an emitting area (A=$785^2$~pc$^2$) that is 10x larger than the CO(6-5) emitting region\footnote{Note that in case of a J-shock the emitting area A required to fit the data would be 1000x larger than the CO(6-5) emitting region. That is why we decided not to show the PDR+J-shock BF model.}. 

The PDR+XDR BF has the lowest $\tilde{\chi}^2_{BF}$ among all the combinations tested in this work ($\tilde{\chi}^2_{BF} = 1.6$), therefore it is our fiducial model. It is composed by a ``cold" and diffuse PDR ($T\simeq 30$ K, ${\rm log}$(n/cm$^{-3}) \simeq 2.5$), that accounts for the CO(1--0) and the CO(2--1) transitions, produced by clouds located at a typical distance $d\simeq500$ pc from the source of the FUV radiation (i.e., OB stars), and by a ``warm" and dense XDR ($T\simeq 65$ K, ${\rm log}$(n/cm$^{-3}) \simeq 4.5$), composed by clouds located at $d\simeq100$ pc from the central X-ray source. In this case, we fixed the XDR column density to a typical value of ${\rm log}$($N_{\rm{H}}$/cm$^{-2})\simeq 23.0$, since X-rays can penetrate deeper into the molecular gas (e.g., \citealt{meije05}).
The radiation penetrates up to a depth $l\approx 5.0$ pc for the PDR, and up to $l \approx 0.3$ pc, for the XDR. These values are in line with the typical environment sizes of giant MCs, and clumps and cores, respectively, as reported by \citet{mckee_ostriker2007}. 
From our fiducial modelling, we estimate a total mass M$_{\rm{gas}} \simeq 3.1 \times 10^9 $ M$_{\odot}$ (typical value of LIRGs, e.g., \citealt{papadopoulos2010}), that appears to be completely dominated by the low-density component that accounts for M~$= (2.9^{+0.9}_{-0.8}) \times 10^9 $ M$_{\odot}$, whereas the warm component contribution is M~$= (2.3^{+0.4}_{-0.5}) \times 10^8 $ M$_{\odot}$. This value is consistent with the total molecular mass found by \citet{fernandez}, who estimated M$_{\rm{gas}}= (2.1\pm0.2) \times 10^9 \, $M$_{\odot}$, considering the CO(1--0) luminosity and the standard conversion factor for starburst systems $\alpha_{\mathrm{CO}}\simeq0.8\, $M$_{\odot}/($K$ \, $km$ \, $s$^{-1} \,$pc$^2)$  \citep{solomon}. From our estimate of the total mass, we infer that in NGC~34 $\alpha_{\mathrm{CO}}\simeq1.1\, $M$_{\odot}/($K$ \, $km$ \, $s$^{-1} \,$pc$^2)$. This value is consistent with what reported by \citet{bolatto} concerning LIRGs. 

\begin{figure}
 \centering
 \includegraphics[width=1.\columnwidth]{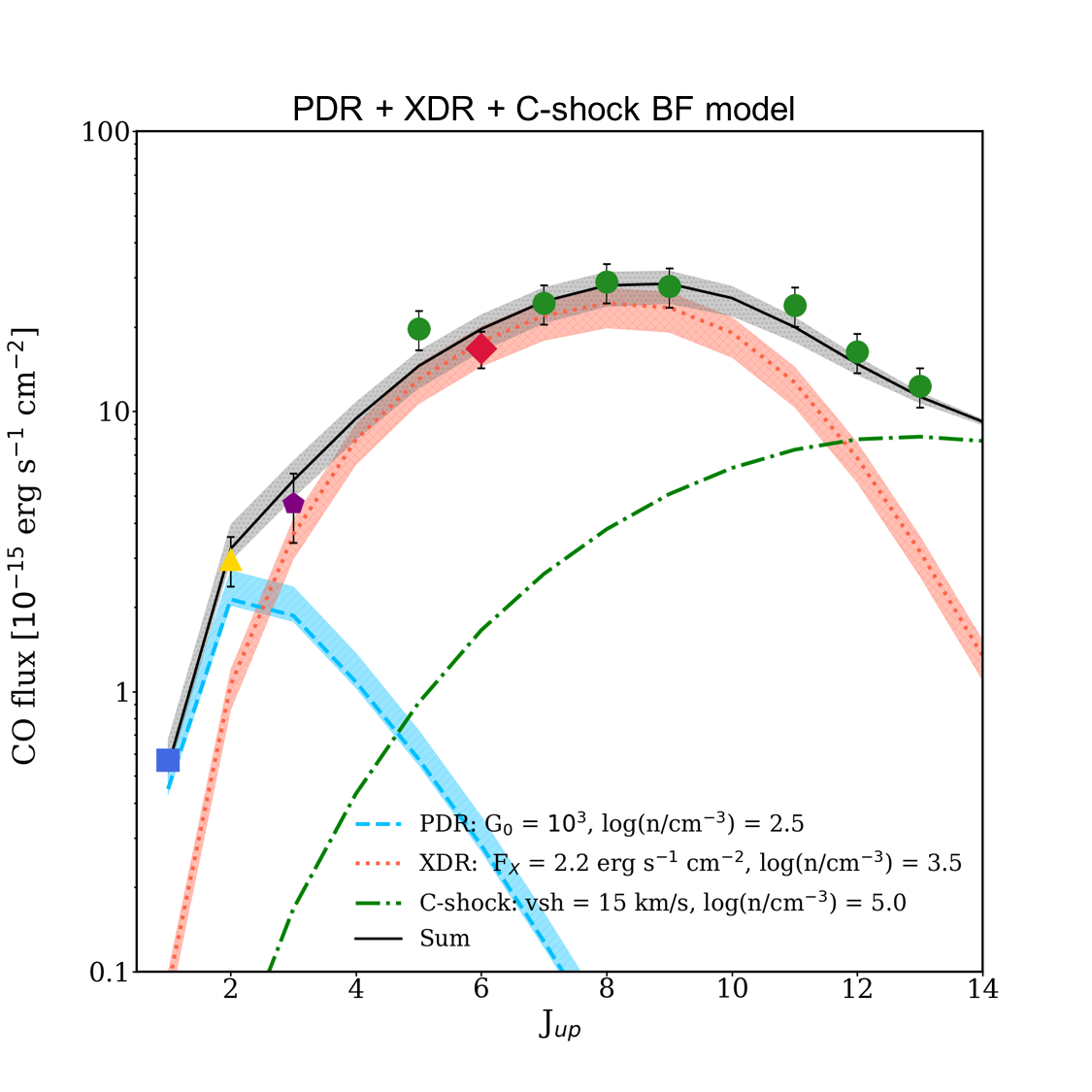}
 \caption{PDR+XDR+C-shock BF model overplotted to the observed data. The light-blue dashed line and the red dotted one indicate the low-density PDR and the XDR, respectively, whereas the green dotted-dashed line represents the C-shock. The solid black line indicates the sum of the three components and the shaded areas indicate the $\pm 1 \sigma$ uncertainty range on the normalisation of each component.}
 \label{fig:best-fit4}
\end{figure}

To complete our analysis, we also explored the possibility of a three-component model accounting for the PDR, XDR, and shock heating contribution. We fixed the column densities of the PDR and the XDR to ${\rm log}$($N_{\rm{H}}$/cm$^{-2}) = 21.8$ and ${\rm log}$($N_{\rm{H}}$/cm$^{-2}) = 23.0$, respectively, and the shock-related emitting area to the CO(6--5) emission size ($\sim 250^2$~pc$^2$). The BF model has a $\tilde{\chi}^2_{BF}= 2.4$ and is shown in Fig.~\ref{fig:best-fit4}, while its parameters are reported in Tab.~\ref{tab:tabbest-fit3comps}. We compare this new configuration with our fiducial PDR+XDR model, taking advantage of the Fisher-test, as presented by \citet{bevington}. We find a value of $f = 0.66 $ that corresponds to a probability $P(F \leq f) = 70$\%, which means that the addition of a third component brings a negligible improvement in the CO SLED fit. This does not mean that shocks are completely absent within the galaxy, but that they do not play a significant role in exciting the high-J CO transitions, more likely powered by the central AGN.

\begin{table}
	\centering
	\caption{BF model (PDR+XDR+C-shock) parameters. For PDRs and XDRs, the values reported are the distance from the source, the density and the column density, respectively. Density and column density values are given in logarithm. For C-shocks, the shock velocity, the pre-shock density and the shock-related emitting area are reported. The * means that the value was fixed in the chi-square analysis. Finally, the ${\chi}^2$ and degrees of freedom (dof) value is indicated.}
	\label{tab:tabbest-fit3comps}
	\begin{tabular}{lccccccccr}  		
		\hline
		\hline
		  Best-fit  	 	  &   d             & n                          & N                            	\\
					  & $[$pc$]$   & $[$cm$^{-3}]$      & $[$cm$^{-2}]$  	   			\\		 
		 \hline
		PDR                   & 500   	     & 2.5    		   &   21.8*		 	      		\vspace{1mm}	\\
		XDR			 & 100	     & 3.5   			   &   23.0*  		     		\\
		\hline
		 		    	 &   v$_{sh}$ & n                          & A                         	               \\
					 & $[$km s$^{-1}]$   & $[$cm$^{-3}]$      & $[$pc$^{2}]$ 			\\	
		\hline
		  C-SHOCK       & 15	     & 5.0   			   &   $($250$)^2$*     			\\
		  \hline
		  \hline
	       $\chi^2$/dof	&	     &			7.1/3	   & 						\\
				         \hline
	\end{tabular}
\end{table} 

\subsection{Discussion and comparison with previous works} 
The CO SLED of NGC 34 has been extensively studied by others authors (e.g., \citealt{rosenberg, pereirangc34}). 

\citet{rosenberg} analysed a sample of LIRGs, including NGC~34, suggesting a qualitative separation on the basis of the shape of the CO ladder, the CO(1--0) linewidth and the AGN contribution, trying to indicate which mechanisms are responsible for gas heating. They identify NGC~34 as a galaxy with a very flat CO ladder beyond the CO(7-6) transition, which indicates a significant emission by warm and dense molecular gas, that is due to another heating mechanism besides UV heating. This galaxy appears to have an inclination corrected CO(1--0) linewidth of $\simeq 600$ km/s and a low AGN contribution of the bolometric luminosity ($\simeq 30$ \%), and thus they infer that the molecular gas in NGC~34 is experiencing shock excitation.

\citet{pereirangc34} have presented a CO emission modelling for NGC~34, using the non-equilibrium radiative transfer code RADEX \citep{radex}, along with \cloudy c13.02 PDR and shock models. They infer that a combination of both PDR and shock models is needed to reproduce the observed CO SLED in NGC~34. They also contend that, although X-ray heating can play a role in AGN, its contribution in LIRGs with high SFR and IR luminosity is negligible in the integrated spectra of the galaxy. 

Our work, on the contrary, suggests that X-rays are important in the CO excitation in NGC 34, even though we do not completely exclude the contribution of shock heating. 
In Sec.~\ref{sez:fitting}, we show that, in order to reproduce the high-J CO emission only with shock models, we need a shock-emitting area far greater than the size of the CO(6--5) area observed with ALMA.
This points towards a scenario in which X-ray heating of high-density gas is favourite against shocks to be the primary source of excitation of high-J CO lines.
Even though it can be rather difficult to distinguish among shock and XDR heating, \citet{meije13} and \citet{gallerani14} suggest that the CO-to-IR continuum ratio ($L_{\mathrm{CO}}/L_{\mathrm{IR}}$) can be a key diagnostic for the presence of shocks. Shocks only heat the gas without affecting the temperature of dust grains, that are the primary source of the IR emission \citep{holl_shocks, meije13} and thus produce high values (L$_{\mathrm{CO}}/$L$_{\mathrm{IR}} >10^{-4}$) for the CO-to-IR continuum ratio. NGC~34 has an observed CO-to-IR continuum ratio of L$_{\mathrm{CO}}/$L$_{\mathrm{IR}} \simeq 2.7\times10^{-5}$, that is almost a factor 4 lower than the threshold. This supports our conclusion that even though shocks are expected to be frequent in the highly supersonic turbulent molecular gas found in LIRGs, they are probably not dominant in the powering of high-J transitions in the observed CO SLED of NGC 34. In this scenario, the large linewidth of the CO(1--0) would not be indicative of shocks, but probably traces the rotation of the $2.1 \,$kpc disk resolved by \citet{fernandez}. We stress that our result is not in contradiction with the relatively low fraction (L$_X/$L$_{bol}\sim 10$~\%) of NGC 34, as in our XDR modelling we have properly scaled our AGN radiation to reproduce the X-ray luminosity of the galaxy.

In the local Universe, a modelling that requires an XDR component to explain the high-J transitions has been proposed also by \citet{vanderwerf} for the nearby ULIRG Mrk 231, and by \citet{pozzi17} for NGC 7130, a nearby  ``ambiguous" LIRG, similar to NGC~34. Both Mrk 231 and NGC 7130 are obscured objects with an intrinsic X-ray luminosity of $\sim 10^{43}$ erg s$^{-1}$, and thus NGC~34, characterised by L$_{\rm{2-10\,keV}} \sim 10^{42}$ erg s$^{-1}$, would extend the importance of an XDR component to galaxies with a lower X-ray luminosity.

\section{Conclusion}\label{sez:4}
In this paper, we analysed archival \xmm, \nustar, ALMA and Herschel data and investigated the molecular CO emission as a function of the rotational level (CO SLED), in order to probe the physical properties of the gas, such as density, temperature and the main source that causes the emission (star formation, AGN or shocks) in the Seyfert 2 and LIRG galaxy NGC~34, comparing PDR, XDR and shock models with the observations. The main steps of our work and our main conclusions can be summarised as follows.
\begin{itemize}
\item The same model has been assumed simultaneously for \xmm \, and \nustar\ data over a broad energy range, fixing the power-law photon indices to $\Gamma=1.9$. We find an obscuration of $N_{\rm{H}}=5.2^{+1.3}_{-1.1}\times10^{23}$ cm$^{-2}$, an observed 2--10~keV flux of 3.2$\times10^{-13}$~\cgs \, and an intrinsic (i.e., corrected for the obscuration) rest-frame 2--10~keV and 1--100~keV luminosities of 1.3$\, \times10^{42}$~\lum\ and 4.0$\, \times \,10^{42}$~\lum, respectively. The 1--100~keV luminosity is used to run Cloudy XDR models.
\item ALMA Cycle 0 data of the CO(6--5) line emission have been analysed starting from the raw data available in the archive. The CO(6--5) integrated flux results to be $(707 \pm 106) \, \mathrm{Jy \, km \, s^{-1}}$ with a peak of  $(213 \pm 32) \,\mathrm{Jy \, km \, s^{-1}}$. The emission comes from a region with a diameter of $\theta \approx1.2$\textacutedbl, that corresponds to a physical scale of $\approx 500$ pc.
\item We infer that a PDR1+PDR2 or a PDR+shock models are unlikely to reproduce the observed CO SLED. The former is discarded because of the high value of the $\tilde{\chi}^2$, while the latter has a BF value of the shock-emitting area that appears to be at least 10x higher than the region associated to the CO(6--5) emission, while shocks are expected to account for the high-J transitions.
\item The observed CO SLED of NGC~34 can be explained by a cold and diffuse PDR ($T\simeq 30$ K, ${\rm log}(n$/cm$^{-3}) \simeq 2.5$, G$_0 = 10^3$), that accounts for the low-J transitions and a warmer and denser XDR ($T\simeq 65$ K, ${\rm log}(n$/cm$^{-3}) \simeq 4.5$, F$_X\simeq 2.2 $ erg$\,$cm$^{-2}\,$s$^{-1}$), necessary to explain the high-J transitions. The existence of a warm XDR component is supported by the $\chi^2$-square analysis and by all the pieces of evidence of a possible AGN in the central region of NGC~34 discussed in Sec.~\ref{sez:ngc34}. We conclude that the AGN contribution is significant in heating the molecular gas in NGC~34.  
 \item The estimated molecular gas mass is M$_{\rm{gas}}\simeq3.1\times10^9$~M$_{\odot}$. We note that the diffuse component contribution to the total mass is dominant compared to the warmer component (one order of magnitude lower). From our total mass estimate, we infer an $\alpha_{\mathrm{CO}}$ of $1.1\, $M$_{\odot}/($K$ \, $km$ \, $s$^{-1} \,$pc$^2)$ for NGC~34.
\item According to the F-test analysis, a three component model composed by a PDR, a XDR and a C-shock is not significantly improved with respect to our fiducial model (PDR+XDR).
 \end{itemize}

Our results shed light on the great potential of combining self-consistent multi-band and multi-resolution data in order to assess the importance of AGN and star formation activity, and mechanical heating produced by shocks for the physics of molecular gas. 
\citet{papadopoulous12}, \citet{imanishi16} proposed that, in addition to CO rotational lines, a combination of low- to mid-J rotational lines of heavy rotor molecules with high critical densities, such as HCO$^+$, HCN, HNC and CN, can be used to probe the large range of physical properties within MCs (T$_{\rm{kin}}\sim15-100 \,$K, n(H$_2)\sim10^2-10^6 \, $cm$^{-3}$, e.g. \citealt{mckee_ostriker2007}). For instance, HCO$^+$ lines appears to be stronger in XDRs than in PDRs by a factor of at least three, while CN/HCN ratio is far higher in PDRs than in XDRs, where it is expected to be $\sim5-10$ \citep{meije07}. 
Therefore, high resolution observations of high critical density molecules with ALMA, characterised by a very high spatial and spectral resolution, could provide new insights on physical properties of NGC~34 and similar nearby composite objects.

\section*{Acknowledgements}
We thank the Italian node of the ALMA Regional Center (ARC) for the support. This paper makes use of the following ALMA data: ADS/JAO.ALMA\#2011.0.00182.S. ALMA is a partnership of ESO (representing its member states), NSF (USA), and NINS (Japan), together with NRC (Canada) and NSC and ASIAA (Taiwan), in cooperation with the Republic of Chile. The Joint ALMA Observatory is operated by ESO, AUI/NRAO, and NAOJ. We kindly thank also G. Zamorani for his useful advice and the anonymous referee for her/his comments and suggestions, which significantly contributed to improving the quality of the publication.

%
%
\bibliographystyle{mnras}
\bibliography{bibliography} 

%
%
%
%

\bsp	
\label{lastpage}
\end{document}